\pdfoutput=1
\documentclass[preprint,aps,superscriptaddress]{revtex4}
\usepackage{graphicx} 
\usepackage{hyperref}

\begin{document}

\title{Metal to insulator transition in epitaxial graphene induced by molecular doping}

\author{S.Y.~Zhou}
\affiliation{Department of Physics, University of California,
Berkeley, CA 94720, USA} \affiliation{Materials Sciences Division,
Lawrence Berkeley National Laboratory, Berkeley, CA 94720, USA}

\author{D.A.~Siegel}
\affiliation{Department of Physics, University of California,
Berkeley, CA 94720, USA} \affiliation{Materials Sciences Division,
Lawrence Berkeley National Laboratory, Berkeley, CA 94720, USA}

\author{A.V.~Fedorov}
\affiliation{Advanced Light Source, Lawrence Berkeley National
Laboratory, Berkeley, California 94720, USA}

\author{A.~Lanzara}
\affiliation{Department of Physics, University of California,
Berkeley, CA 94720, USA} \affiliation{Materials Sciences Division,
Lawrence Berkeley National Laboratory, Berkeley, CA 94720, USA}

\date{\today}


\begin{abstract}
The capability to control the type and amount of charge carriers in a material and, in the extreme case, the transition from metal to insulator is one of the key challenges of modern electronics.  By employing angle resolved photoemission spectroscopy (ARPES) we find that a reversible metal to insulator transition and a fine tuning of the charge carriers from electrons to holes can be achieved in epitaxial bilayer and single layer graphene by molecular doping.  The effects of electron screening and disorder are also discussed.  These results demonstrate that epitaxial graphene is suitable for electronics applications, as well as provide new opportunities for studying the hole doping regime of the Dirac cone in graphene. 
\end{abstract}

\maketitle

Chemical doping is a well-known and widely used method of manipulating the electronic properties of materials.  For example, it can convert an insulator into a high temperature superconductor or into a material with colossal magnetoresistance.  The modern microelectronics industry probably provides the most notable example since it is almost entirely based on doped semiconductors.  Finding a new material suitable for microelectronics is not easy since one has to find how to change its carrier concentration via doping.  In addition, in the ultimate case, it might be necessary or at least useful to induce a reversible metal to insulator transition (MIT) in a material.  In the past few years, a great deal of attention has been focused on graphene - a few layers thick graphite sheets, which is considered by many as a material of choice for the future \cite{GeimRev, ANCHRMP}.  Although it would certainly be a major breakthrough to induce a MIT in graphene, so far this is reported only in mechanically exfoliated graphene by using a complicate gate device configuration \cite{GapBilayer, Gap2MLGraphene} which is difficult for large scale fabrication.  
Recently it has been shown that, when graphene is epitaxially grown on a Si-terminated semiconducting substrate, a gap opens at the Dirac point both for single layer \cite{NatMat} and bilayer graphene \cite{Eli,NatMat}.  However, because of the charge transfer from the substrate \cite{Liz, Eli, NatMat} the Dirac point and hence the gap, lie well below the Fermi energy $E_F$.  Therefore, although epitaxial graphene could be an ideal system to observe a MIT, this can only be realized if E$_F$ moves into the gap region.  Unfortunately, the control of charge carriers in epitaxial graphene is not as simple as for exfoliated graphene: doping by applying a gate voltage to change charger carriers from electrons to holes has been achieved only recently, and only for a small doping range \cite{Kedzierski},
and {\it hole} doping by chemical methods has not yet been demonstrated successfully.

\begin{figure}
\includegraphics[width=5.8cm] {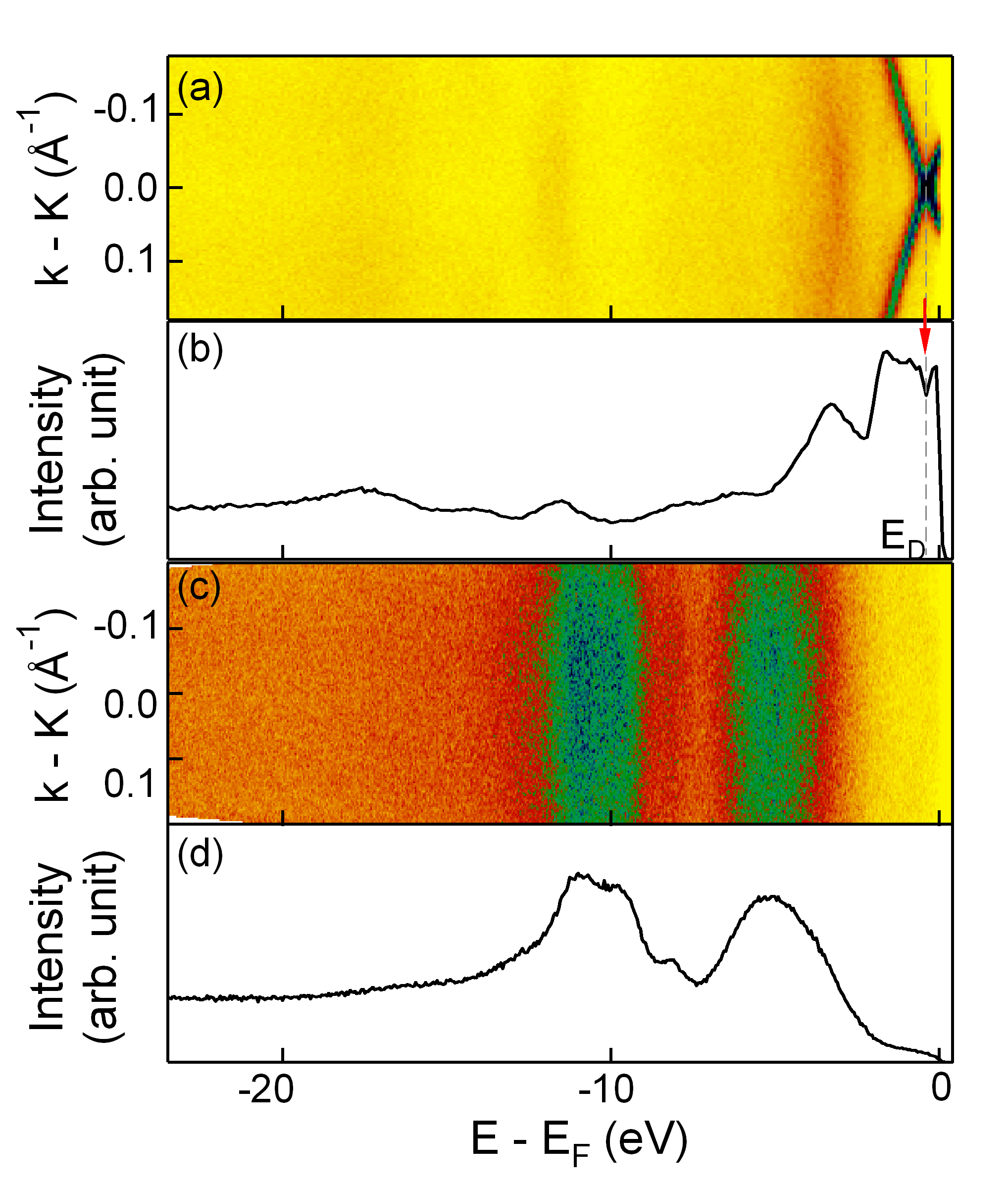}
\label{Figure 1} \caption{Dispersions through the K point for the as-grown single layer graphene (a) and for the sample with the highest NO$_2$ doping (c). Panels (b) and (d) show the angle integrated spectra for (a) and (c) respectively. In panels (c) and (d) note the appearance of the NO$_2$ states at $\approx$ 5 and 11 eV below E$_F$.}
\label{NO2vb}
\end{figure}

\begin{figure*}
\includegraphics[width=14. cm] {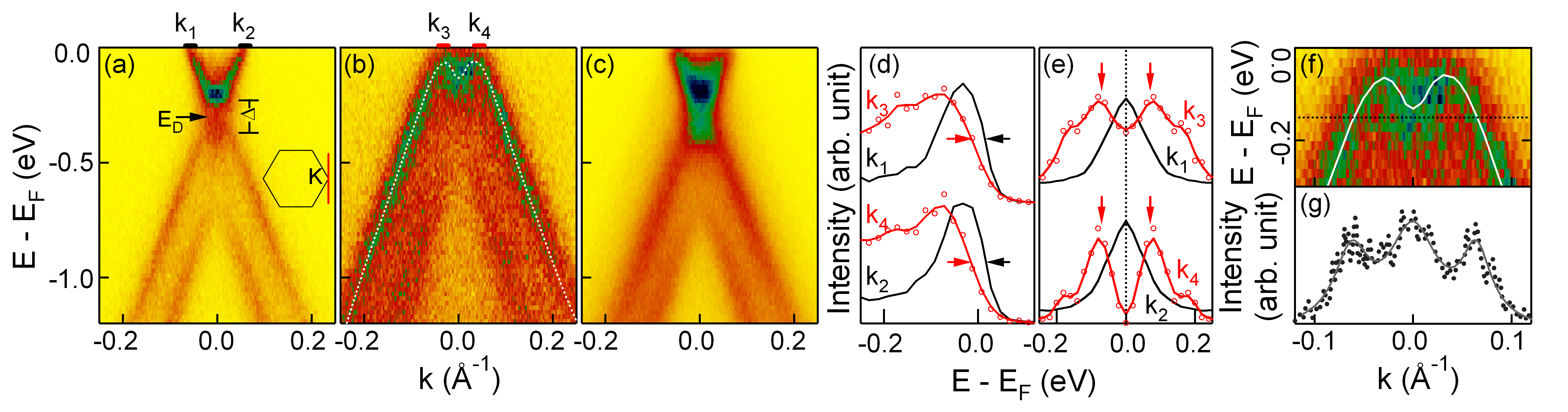}
\caption{(a) Dispersions of as-grown bilayer graphene for a cut through the K point (see red line in the inset). (b) Dispersions taken after 0.6 Langmuir (1 Langmuir=$10^6 torr\cdot s$) NO$_2$ adsorption. (c) Data taken after NO$_2$ desorption by synchrotron beam. (d) EDCs at k$_1$, k$_2$, k$_3$ and k$_4$ as labeled on the top of panels a and b. The red dots are the raw data points and the red lines are the smooth data as a guide for the eye.  The black and red arrows point to the midpoint of the leading edge, which is shifted to higher binding energy after NO$_2$ doping.  (e) Symmetrized EDC with respect to E$_F$.  
The absence of a peak at E$_F$ in the red curves shows the insulating property of the sample in panel b.  (f) Zoom-in of data shown in panel b. The white line is a guide to the eye for the dispersions. (g) Momentum distribution curve (MDC) at the energy labeled by the dotted black line in panel (f). The dots are the raw data and the solid line is the fit using three Lorentzian peaks simulating the cross-section of the hat-like dispersion in panel e.}
\label{NO2bilayer}
\end{figure*}

Motivated by the reports that molecular adsorption of NO$_2$ can induce hole doping in carbon nanotubes \cite{KongSci} and exfoliated graphene \cite{GeimNatMat, Wehling}, we have investigated the effect of such doping on the electronic structure of single layer and bilayer epitaxial graphene by using angle-resolved photoemission spectroscopy (ARPES).  ARPES is a powerful tool for studying the doping effects and metal insulator transition \cite{Perfetti} as it 
directly detects the excitation gap, which distinguishes an insulating behavior from a metallic behavior.  Here we report a reversible metal to insulator transition associated with almost rigid shift of the band structure as a function of doping, leading to a doping independent Fermi velocity and electron-phonon coupling.  Moreover we found that scattering by charge impurities induced by NO$_2$ adsorption is suppressed by electron screening, in line with the report of constant mobility by transport measurements \cite{GeimNatMat}.  These results mark an important step on a route to integrating graphene into semiconductor technology.

Single layer and bilayer graphene samples were grown {\it in situ} on the silicon-terminated face of the SiC substrate by thermal decomposition of electron doped SiC wafers as described elsewhere \cite{BergerJPC, Liz}.  The average domain size for the monolayer graphene is larger than 50 nm \cite{ZhouReply}.  Adsorption of NO$_2$ molecules on otherwise clean graphene samples was achieved via controlled exposures to the NO$_2$ gas (Matheson, 99.5$\%$). All ARPES spectra were taken at beamline 12.0.1 of the Advanced Light Source (ALS) in Lawrence Berkeley National Laboratory using SES100 electron spectrometer. The photon energy was 50 eV and the energy resolution was better than 30 meV.  The measurement temperature was 20 K, well below the threshold (140K) for thermal desorption of NO$_2$ from graphite \cite{NO2graphite}. We studied NO$_2$ adsorption in two ways: ARPES data were either taken under the continuous flow of NO$_2$ at a pressure of $\approx$ 5$\times$10$^{-8}$ torr or by exposing the sample to the gas for a certain time and taking data after the chamber was pumped down to the base pressure. Both measurements yielded identical results.  Although different doping levels were studied, an exact estimation of the amount of NO$_2$ adsorbed on the surface of our samples is difficult due to the significant photo-stimulated desorption. To minimize the impact of the photo desorption on our data we had to intentionally reduce the photon flux and accumulate the spectra for only 30 seconds.  This explains a relatively poor statistics of the data taken from graphene with adsorbed NO$_2$.

Before discussing the effect of molecular doping on the electronic structure of bilayer and single layer graphene, we shall prove that NO$_2$ is present on the surface of our samples. Figure \ref{NO2vb} shows ARPES data of the valence band of a single layer graphene sample before (panel a) and after (panel b) exposure to NO$_2$.  The exposure to NO$_2$ leads to the appearance of two broad peaks centered at 5 eV and 11 eV below E$_F$, similar to those reported previously on the surfaces of W(110) \cite{NO2onW} and Si \cite{NO2onSi}.  Previous studies of NO$_2$ adsorption on W(110) \cite{NO2onW} and graphite \cite{NO2graphite} suggested that NO$_2$ forms dimmers (N$_2$O$_4$) upon adsorption.  A similar dimmerization might also occur here.  This is supported by the absence of any spin-polarized signal in the density of states for any doping value (not shown), predicted in the case of NO$_2$ adsorption but not in the case of dimers (N$_2$O$_4$) \cite{Wehling}.

\begin{figure*}
\includegraphics[width=11.6 cm] {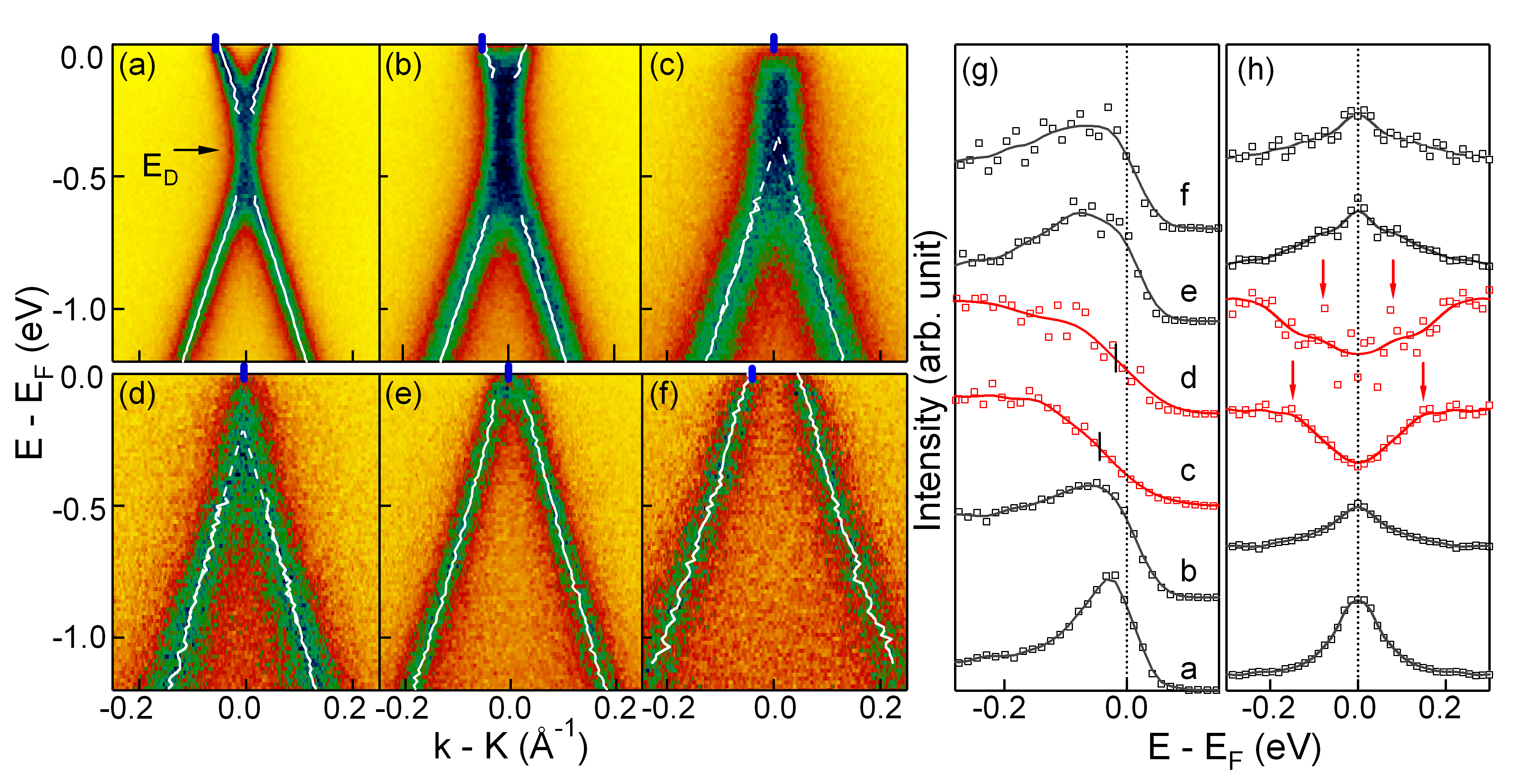}
\label{Figure 3} \caption{(a-f) Dispersions for single layer graphene through the K point, for the as-grown sample (a) and for various dopings after NO$_2$ adsorption (b-f). The white lines are the dispersions extracted from fitting the MDC peaks. When the MDC peaks cannot be clearly resolved the extrapolated dispersion is shown (see dotted lines in panels c-d). (g) EDCs taken in the momentum regions (indicated by a small tick mark on top of each panel) where the bands are closest to E$_F$.  The symbols are the raw data and the solid lines are guides for the eye. (h) Symmtrized EDCs for data shown in panel g with respect to E$_F$.} 
\label{NO21ML}
\end{figure*}

Fig.~\ref{NO2bilayer} shows ARPES data taken on bilayer graphene before and after NO$_2$ adsorption at 20K. Panel a shows dispersions through the K point in the as-grown graphene. The Dirac point energy is readily identified at the binding energy of 0.3 eV.  Strictly speaking, the Dirac point, which separates the upward dispersing conduction band from downward dispersing valence band, is located within a gap. Existence of this gap between valence and conduction bands in bilayer graphene has already been shown in previous ARPES studies \cite{Eli, NatMat}.  Panel b shows data taken right after dosing NO$_2$ gas at 1$\times$10$^{-8}$ torr for 60 seconds (0.6 Langmuir of NO$_2$ gas). Clearly, adsorption of NO$_2$ leads to the shift of the entire band structure toward E$_F$, which indicates hole doping of the bilayer graphene.  Close inspection of the region near E$_F$ shows that E$_F$ lies within the gap separating the conduction and valence bands with the latter band located above E$_F$.  To illustrate this, in Fig.~\ref{NO2bilayer}(d) we plot the energy distribution curves (EDCs) at the momentum values where the bands are closest to E$_F$ for data shown in panel a (black curves) and panel b (red curves).  It is obvious that after NO$_2$ adsorption, there is a depletion of the states at E$_F$, accompanied by a shift of the leading edge to higher binding energy by $\approx$ 30 meV. The absence of spectral weight at E$_F$ is the typical characteristic of a gap, whose magnitude is measured in general from the mipoint in the EDC spectra at k$_F$ (leading edge) \cite{ShenPRL}.  The data shown in panel (d) hence supports the insulating nature of our sample upon NO$_2$ absorption (panel b), with a gap of $\approx$ 30 meV.  
In panel e we show the EDC spectra symmetrized at E$_F$. This method allows a direct visualization of the gap by eliminating the contribution of the Fermi function from the ARPES data, as has been widely used in the case of high T$_c$ superconductors \cite{Norman}.
The presence of the peaks at E$_F$ in the symmetrized EDCs in the as-grown sample and the suppression of intensity at E$_F$ after NO$_2$ adsorption show that the sample undergoes a metal insulator transition upon NO$_2$ adsorption.  This metal insulator transition is reversible: the sample becomes a metal again if we remove most of the NO$_2$ from the surface by exposure to high photon flux, or by annealing the sample. Fig.~\ref{NO2bilayer}(c) displays the data taken after desorption of NO$_2$ by photon flux. The bands of graphene are shifted back to their original positions with the Dirac point at 0.3 eV below E$_F$. However, the peaks in panel 2(c) are still broader than those in panel 2(a), which might indicate a small fraction of residual NO2 molecules.  By thermal annealing at $\approx$ 400$^\circ$C the residual molecules can be completely removed, as confirmed by the sharpening of the peaks and the complete recovery of the line width of pure graphene.

In addition to demonstrating molecular hole doping of graphene, the data in Figs.\ref{NO2bilayer}(b,f) also allow an unobscure view of the top of the valence band which exhibits a characteristic hat-like shape \cite{Eli, McCannPRL}.  The persistence of this hat-like shape upon NO$_2$ doping is important, as it has been predicted that with increasing amount of disorder, the dispersions near E$_D$ will become broader and blurry significantly \cite{ImpurityBilayer, NilssonDisorder}.  The observation of the hat-like structure in the data suggests that electron screening in bilayer graphene is very effective to suppress the scattering by charged impurities induced by NO$_2$ adsorption.  This is in line with transport measurements on exfoliated graphene \cite{GeimNatMat}, where the mobility is almost unchanged upon NO$_2$ adsorption.  

Finally, we wish to attempt to estimate the amount of NO$_2$ adsorbed on the graphene.  The amount of carrier concentration in panel a has been estimated by measuring the area enclosed by the Fermi contours and through the Luttinger theorem.  Assuming that the Fermi surface is a circle, the charge concentration in panel a is 0.0052 electrons per unit cell.  Upon exposing the graphene sample to 0.6 Langmuir of NO$_2$, the sample becomes insulating and the Fermi surface disappears. If we assume that each NO$_2$ molecule accepts as much as 1 electron from graphene as suggested by transport measurement \cite{GeimNatMat}, change of 0.0052 electrons per unit cell corresponds to 0.033 Langmuir of NO$_2$.  This suggests that with the dosage 0.6 Langmuir NO$_2$ molecules, only 5.5$\%$ of the molecules stick to graphene and remove 1 electron from graphene.  This is reasonable considering the fast photo-desorption of the NO$_2$ molecules.

\begin{figure}
\includegraphics[width=8.0 cm] {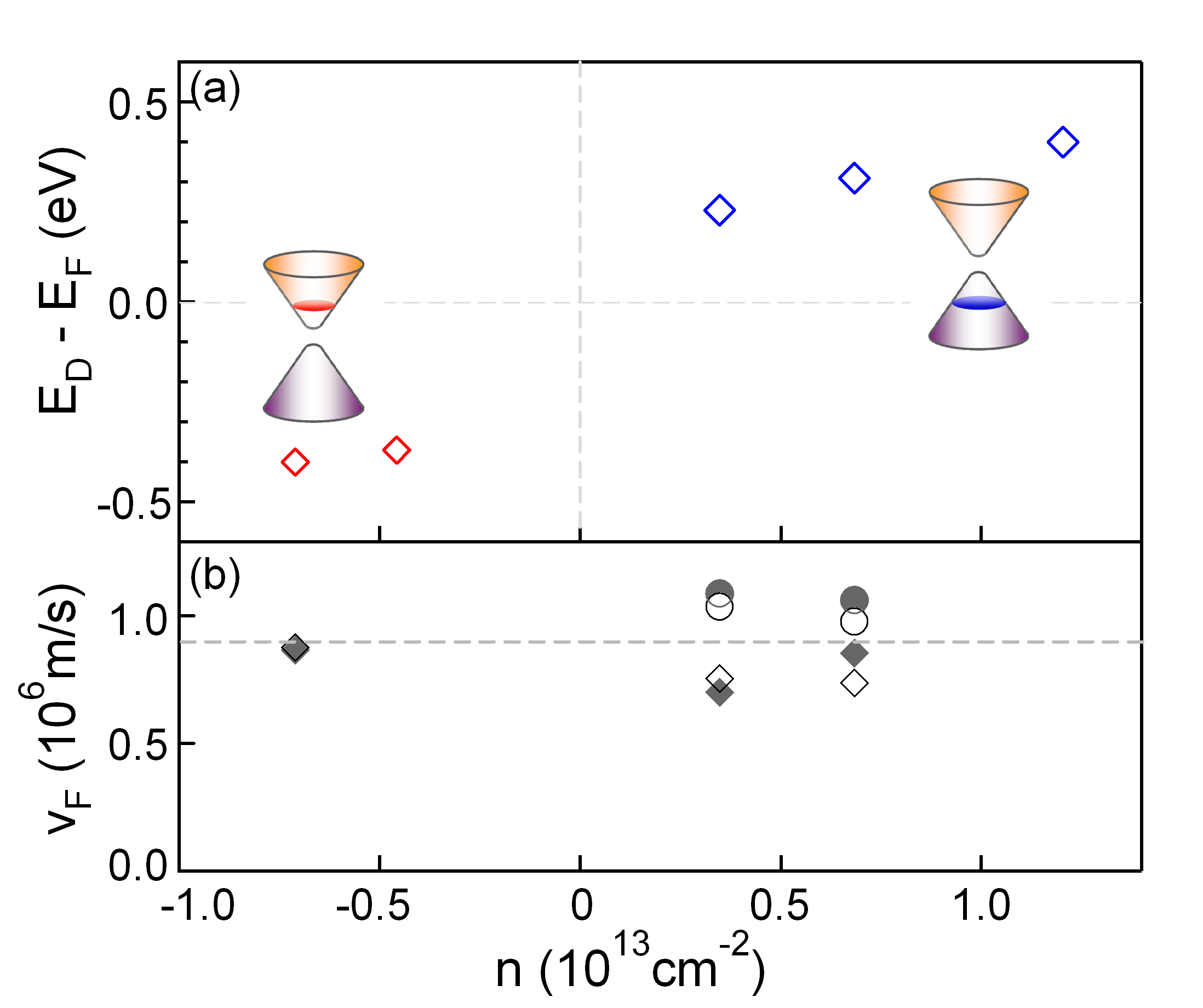}
\label{Figure 4} \caption{(a, b) Plot of E$_D$ and Fermi velocity as a function of the carrier concentration for data taken on single layer epitaxial graphene. In panel b, the data are extracted by fitting the dispersion between E$_F$ and -0.3 eV (circles) and between E$_F$ and -0.1 eV (diamonds). The open symbols are extracted from the dispersions on the left and the filled symbols from the dispersions on the right.}
\label{NO2doping}
\end{figure}
  
In Figure \ref{NO21ML} we show the effect of molecular doping on single layer epitaxial graphene for progressive adsorption of NO$_2$ (panels a-f).  Data in panel a are taken on the as-grown sample, which is electron doped with the Dirac point located at 0.4 eV below E$_F$.  As previously reported \cite{NatMat, EliNatPhys} the valence and conduction bands deviate from the expected conical dispersion near the K point, where an anomalous extended vertical region of finite intensity is observed.
The origin of this deviation is under debate \cite{NatMat, EliNatPhys}.  
By hole doping we have the unique ability to directly access this region and to investigate its origin.
Similar to the case of bilayer graphene, adsorption of NO$_2$ leads to 
a significant shift of the entire band structure.  
More specifically the chemical potential can be shifted from the conduction band (panels a-b) all the way down to the valence band (panels e-f), with a maximum shift of 0.8 eV in energy and a total change of 1.9$\times$10$^{13}$ cm$^{-2}$ in carrier concentration.  The latter corresponds to applying a gate voltage as large as 260 V in exfoliated graphene \cite{NovoselovSci, NovoselovPNAS, ZhangNat}.  
When the chemical potential falls in the anomalous region near E$_D$ (panels c and d), extrapolation of the valence band shows that the top of this band lies below E$_F$ (see white dashed line) while the bottom of the conduction band lies above E$_F$, the two being separated by a region of finite intensity pinned at the K point. 
Moreover, the EDCs at the K point show a depletion of spectral weight, the absence of a Fermi cutoff and a clear shift of the leading edge towards higher binding energy (see curves c and d in panel g).  
This is in contrast to the case in which the chemical potential falls within the conduction or valence bands (curves a, b, e, f of panel g) where a peak near E$_F$ (dotted line) and a Fermi crossing are observed. 
This shift of the leading edge resembles that of bilayer graphene discussed above when the chemical potential falls within the gap region (Fig.~\ref{NO2bilayer}(d)), and therefore, is suggestive of an insulating state even for the single layer graphene, which is similarly reversible by exposure of the surface to high photon flux or by annealing the sample at high temperature.   Moreover, this result confirms that the gap between the conduction and valence band is indeed an intrinsic property of epitaxial graphene as previously argued \cite{ZhouReply}, as is also supported by recent ab initio calculation \cite{Kim} and the absence of the gap in the preliminary ARPES study of exfoliated graphene \cite{Knox}.

Figure \ref{NO2doping} summarizes the effect of NO$_2$ adsorption in single layer graphene. Fig.~\ref{NO2doping}(a) shows the amount of hole doping estimated from the shift of E$_D$ (vertical axis) and the change of charge carrier concentration (horizontal axis).  Upon NO$_2$ doping, the Fermi energy can be shifted by as much as 0.8 eV, and the charge carriers can be tuned in a wide range from 0.7$\times$10$^{13}$ cm$^{-2}$ electrons to 1.2$\times$10$^{13}$ cm$^{-2}$ holes.  Fig.~\ref{NO2doping}(b) plots the Fermi velocity v$_F$=$\frac{1}{\hbar}\frac{\partial{E}}{\partial{k}}$, a quantity that governs the low-energy quasiparticle dynamics, as a function of carrier concentration.  Surprisingly we find that the Fermi velocity is nearly constant for all doping levels to within $\pm$20$\%$ of the initial value, though small changes, expected on a scale which is beyond the resolution of the present experiment \cite{Louie}, cannot be excluded.  This independence of the Fermi velocity points to a doping independent electron-phonon coupling, in contrast to an earlier report where a huge change of the Fermi velocity by electron doping through K and Ca \cite{Jessica} was reported. Whether this suggests the presence of a strong electron-hole asymmetry or the hybridization of graphene $\pi$ bands with those from K and Ca \cite{NetoK, Calandra, ParkPRB} in the previous case \cite{Jessica}, has to be confirmed.  Finally, we note that a similar doping independent Fermi velocity has been reported for the nodal quasiparticles in high temperature superconductors, which are also Dirac fermions, through the insulator to superconducting transition \cite{XJZhou}.  Whether this universality is a general property of Dirac Fermions is certainly an interesting topic that needs to be further investigated.

In summary, we have performed ARPES measurements of the electronic structure of single and bilayer epitaxial graphene with adsorbed NO$_2$.  Our data directly prove that NO$_2$ induces hole doping of graphene over a wide doping range, tuning the charge carriers from electrons to holes. This results in: a rigid shift of the band structure, and hence a doping independent Fermi velocity and electron-phonon coupling; a reversible MIT; and, in the case of bilayer graphene, a suppression of scattering by charge impurities by electron screening.  Our findings provide new directions for achieving semiconducting graphene and open up an interesting route to studying the physics of the hole doping regime of the Dirac cone.

We thank B.S. Mun, B. Freelon for providing us the NO$_2$ gas, A.H. Castro Neto, D.-H. Lee and M.I. Katsnelson for useful discussions.  The sample growth and ARPES measurements work were supported by the National Science Foundation through Grant No.~DMR03-49361.  The ARPES measurements and data analysis were supported by the Division of Materials Sciences and Engineering of the U.S Department of Energy under Contract No.~DEAC03-76SF00098.  S.Y. Zhou thanks the Advanced Light Source Doctoral Fellowship for financial support.

\end{document}